# 10 Inventions on keyboard key switch mechanism
# -A study based on US patents

**Umakant Mishra**

Bangalore, India

umakant@trizsite.tk

http://umakant.trizsite.tk

**Contents**

# 1. Introduction

A keyboard the main input device for a computer. A standard computer keyboard comprises of a base stand, a membrane circuit board with pressure sensors and several keys with a key pressing mechanism.

A standard keyboard has four sections on the keyboard, a text entry section, a navigation section, a numeric section and a function key section. Each section comprises of several keys.

The number of keys may vary from keyboard to keyboard. A standard present day keyboard contains 101 keys. Some older keyboards were comprised of 81 keys. Laptop and portable computers may have less number of keys. Some special keyboards may have more number of keys.

**The keyboard key switch mechanism**
The key switches of the keyboard of a computer system are generally comprised of a key cap having a plunger, conductive rubber disposed above a membrane circuit and compressed by the plunger to trigger the membrane circuit causing it to produce an electric signal to the computer.

Some key switches use springs. Some other keyboards use rubber domes or a dome sheet, which do the function of springs. When the user depresses the key button the spring or domes collapse. The key switch depresses the key stem, which actuates the button on the membrane circuit. When the user releases the button the springs or rubber domes push the button up to the rest position.

# 2. Inventions on key switch mechanism

### 2.1 Computer keyboard with flexible dome switch layer (Patent 5212356)

**Background problem**
Many computer keyboards use a flexible sheet of nonconductive material beneath the key caps in which the flexible sheet has molded dome portions at each key position to serve as a return spring.
Although this mechanism is popular, it has some drawbacks. It is expensive and time consuming to manufacture. Secondly, a separate mold is required to be constructed for each different key layout.

**Solution provided by the patent**
George English disclosed a solution to the above problem (Patent 5212356, assignee- Key Tronic Corporation, issued May 1993). The invention uses parallel extruded ridges at desired intervals for each key row. The ridge domes have parallel front and real walls that are collapsible to give the spring effect to lift up the depressed switches. The bottom surface of the bridging crown has an actuation keel of a conductive material, which actuates the keyswitch when the key is depressed.

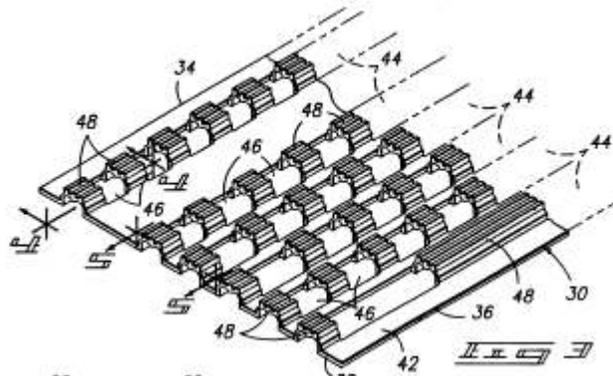

This invention has the advantage that the same rubber sheet can be used with different layout of keyboards.

**TRIZ based analysis**

It is desirable to use the same rubber dome sheet for different keyboard layouts, instead of manufacturing different rubber domes for each minor change in the keyboard layout **(Desired result)**.

As the invention uses parallel ridges for each row of the keyboard that can be used with different structure of keyboard **(Principle-6: Universality)**.

The invention uses ridges (for each row of keys) instead of pointed domes (for each individual key) **(Principle-17: Another dimension)**.

**2.2 Key structure for computer keyboards (Patent 5372442)**

**Background problem**

When the user depresses a button on the keyboard, the keyboard generates a character based on the input of the user. But sometimes because of improper structure of the keyboard, the character is not generated even if the user depresses a key. It is necessary to improve the key pressing mechanism for reliable performance.

**Solution provided by this invention**

Kun-Chu Wang disclosed a structure of keyboard (Patent 5372442, assignee-Nil, issued Dec 1994) having key units that includes (i) a hollow housing securely mounted on a base frame of a computer keyboard, (ii) an actuating member and (iii) a spring member.

The actuating member has two parts, the first part includes a distal activating end for impinging on the base frame and the second part is elastic and includes a bent section and a distal actuating end.

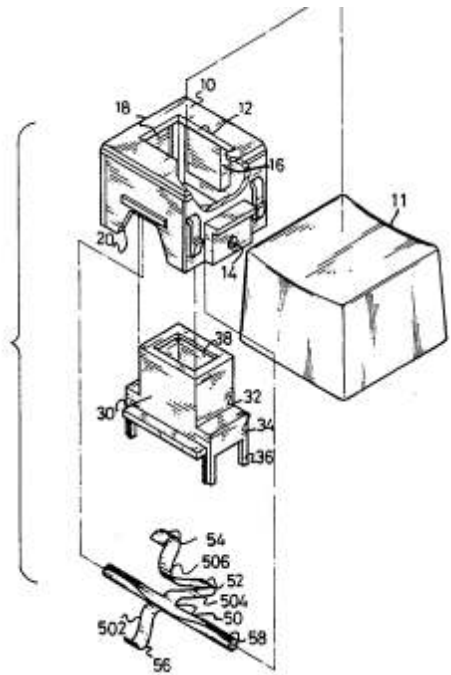

When the user presses the key, the actuating member moves downward and impinges on the contact to send a specific signal, while the bent section impinges on the resounding plate that generates a click.

### TRIZ based analysis

The mechanical section of the key is divided into different sections, viz., a key, a hollow housing, the actuating member, the spring etc. **(Principle-1: Segmentation)**.

Each keystroke generates a click sound that ensures the reliability of the operation **(Principle-23: Feedback)**.

### 2.3 Computer keyboard with integral dome sheet and support pegs (Patent 5430263)

### Background problem

There are some drawbacks of a traditional keyboard. The traditional keyboards are expensive to manufacture. There is an increasing drive among keyboard manufacturers to reduce the cost of manufacturing.

Besides, the conventional keyboard increases friction of the keys with the age of the keyboard and the keys often "stick" in the depressed position or return very slowly to the rest position. This causes the user to strike harder which results in user fatigue and further degradation of the keyboard.

The other drawback of some keyboards is that the separate keys must be separately and independently mounted in their corresponding key supports, which requires significant amount of time.

**Solution provided by the invention**

English et al. invented a keyboard (Patent 5430263, assignee- Key Tronic Corporation, issued- July 1995) in which the individual rows of keys are integrally formed with a common base unit or mounting strip. The function keys, other individual keys are also mounted to the strip. The lows of cantilevered keys are arranged to partially overlap adjacent rows such that the keys in one row actuate switch contacts aligned beneath mounting strips in the adjacent row. The keyboard has a switch contact membrane, a printed circuit board and a dome sheet.

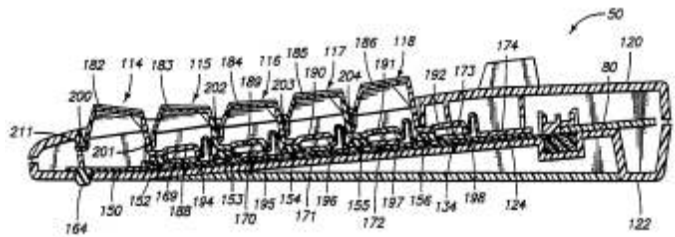

**TRIZ based analysis**

The invention eliminates the conventional key stem and use a dome sheet made of an elastomeric material or rubber **(Principle-28: Mechanics substitution, Principle-30: Thin and flexible)**.

**2.4 Computer keyboard key switch (Patent 5457297)**

**Background problem**

The conventional key switches use a spring inside the key in order to bring the key cap back to the return position. But the use of the spring does not permit the height of the key switch to be reduced.

There is another disadvantage of using a spring. While operating the keyboard, if the user touches the border or any corner of the key cap, the key cap may not move the plunger down which causes a key-in error.

**Solution provided by the invention**

Pao-Chin Chen invented a switch (Patent 5457297, issued Oct 1995), which eliminates the above problems. According to the invention the keyboard key switch includes a bottom support board, a membrane circuit supported on the support board, a key base having a rubber cone and supported on the membrane circuit, a bridging device supporting board supported on the key base, a key cap, and a bridging device connected between the key cap and the bridge device supporting board and consisting of two rectangular open frames pivotally connected into a crossed form for permitting the key cap to be depressed to compress the rubber cone causing it to trigger the membrane circuit.

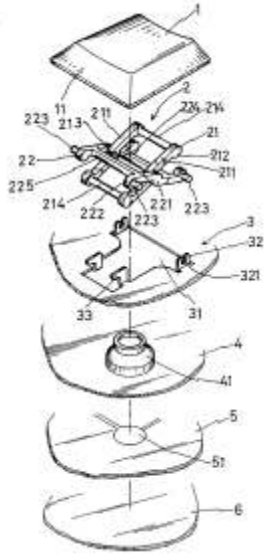

**TRIZ based analysis**

The keyboard uses a bridging device and a rubber cone instead of a spring to trigger the membrane circuit **(Principle-28: Mechanics Substitution, Principle-30: Think and Flexible)**.

**2.5 Computer keyboard with cantilever switch and actuator design (Patent 5481074)**

**Background problem**

One of the drawbacks of the prior art keyboard is the bearing interface between monoblock key support and key stem of keys. As the age of the keyboard, the key stem creates surface friction while sliding within the key support, and begins to move less freely. This causes the users to increase their pushing force and ultimately the keys to "stick" in the depressed position.

**Solution provided by the invention**

George English invented a computer keyboard (Patent no 5481074, assignee: Key Tronic Corporation, issued Jan 1996) with multiple rows of cantilevered keys which are flexibly attached to first common mounting strips. The keyboard also has multiple rows of cantilevered sub-members flexibly attached to second common mounting strips, with the sub-members being aligned beneath associated keys. The cantilevered sub-members are in sliding contact with their associated cantilevered keys and induce a tactile "break over" sensation as the associated cantilevered keys are depressed. Each sub-member is designed to actuate a switch contact as the cantilevered key is depressed, whereby the computer keyboard can be constructed without a dome sheet.

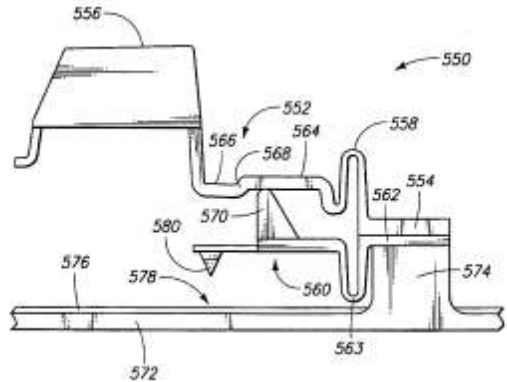

**TRIZ based analysis**

The invention uses cantilevered keys instead of a dome sheet to actuate the keys **(Principle-28: Mechanics substitution)**.

**2.6 Rubber cone layer of a keyboard (Patent 5760351)**

**Background problem**

Normally the keyboard contains a rubber cone layer which is mounted between the frame and membrane circuit. When a key switch is depressed, the corresponding rubber cone is compressed and forced downwards to trigger the membrane circuits causing it to produce a respective electrical signal. When the key switch is released, the corresponding rubber cone returns to its former shape to push the key switch back to its former position.

Because the bottom side of the rubber cone layer and the top side of the membrane circuit are smooth surfaces, a vacuum tends to be produced between the rubber cone layer and the membrane circuit causing the rubber cone layer and the membrane circuit to be stuck together. How to solve this problem?

**Solution provided by the invention**

Ching-Cheng Tsai invented a rubber cone layer (Patent 5760351, assigned to Chicony Electronics, Issued in June 98) having fine grains distributed over the top and bottom sides that prevents sticking to the membrane circuit of the computer keyboard. This method also facilitates its removal from the mould during its fabrication.

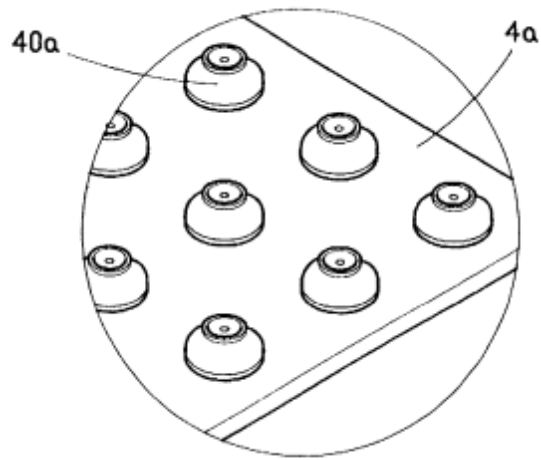

**TRIZ based analysis**

The rubber cone layer and membrane circuit should not get stuck together **(desired result)**.

The invention spreads fine grains on both the sides of the membrane that allows air the pass through without causing a vacuum to be formed **(Principle-3: Local quality, Principle-31: Hole)**

**2.7 Plane mechanical keyboard (Patent 5861588)**

**Background problem**

There are different types of mechanical keyboards available in the market. They include membrane-type keyboards, flexible contact keyboards and also touch pad keyboards. But the pocket keyboards are very small for the human fingers to operate comfortably for long time. They need great precision in pressing the key while typing. Because of the miniature size, the user sometimes presses the keys adjacent to the desired key resulting in multiple key depressions. There is a need to solve this problem.

**Solution provided by the invention**

Gillot invented a plane mechanical keyboard (Patent 5861588, assigned to France Telecom, Issued in Jan 1999) that comprises several main keys for alphanumeric characters, each of which is surrounded by some secondary keys. The mechanism is done in such a way that when the user presses the main key, the neighboring secondary keys are also pressed downward. This creates a pull back force on the neighboring main keys to pull upwards. This pull back action to the adjacent keys prevents them being pressed.

This simple mechanism reduces the necessary precision of striking and eliminates the possibility of more than one key being pressed at a time as the neighboring keys around the pressed key cannot be pressed as they are pulled up. This keyboard is found suitable for pocket computers, portable telephones and similar small equipments.

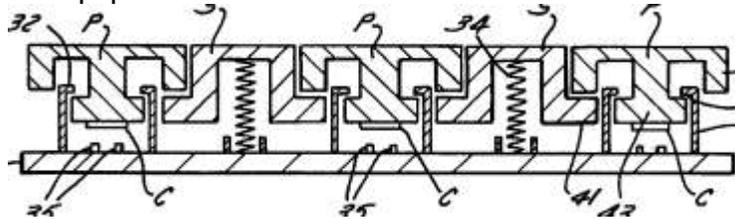

### TRIZ based analysis

The computer should not accept input from any key other than the intended key **(Ideal Final Result)**. The pressing action of a key itself should prevent depression of adjacent keys **(desired result)**.

We need small keys and less space between keys to accommodate large number of keys in a small portable keyboard. But reducing the size of the keys and space between keys leads to depression of undesired adjacent keys. Thus we want small keys but don't want to press undesired adjacent keys, which causes a contradiction between the size and precision or accuracy **(Contradiction)**.

The invention builds the keys in such a way that depression of one key does not allow depression of adjacent keys simultaneously **(Principle-8: Counterweight)**.

### 2.8 Keyboard with adjustable height (Patent 5874696)

#### Background problem
It is nice to have a normal operating height of the key switches for typing comfort. But the height of the keys occupies vertical space during storage. It's necessary to reduce the height of the keys to a more compact size when the keyboard is not used.

#### Solution provided by the invention
Hayashi, et al. invented a method (Us patent 5874696, assigned to Fujitsu Takamisawa Component Ltd, Feb 99) to allow the height of the key switches on the keyboard to be lowered while closing the cover and automatically reinstated to the normal operating height when the cover is opened. This result is achieved by placing an elastic member between the keyboard base and the keys. The keyboard is designed to move up and down by depressing the elastic plate.

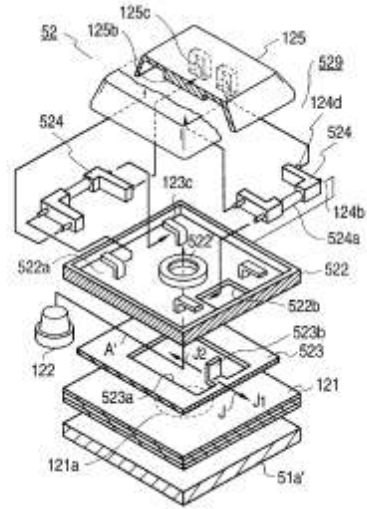

**TRIZ based analysis**

The height of the keys should be more while used, and less while closed inside the laptop box **(Contradiction)**.

The invention uses an elastic member between the keyboard and the keys which is expanded during use and depressed while closed **(Principle-24: Intermediary, Principle-15: Dynamize)**.

**2.9 Computer keyboard with adjustable force keystroke feature using air pressure (Patent 5879088)**

**Background problem**

Some users typically press hard and some user press soft. The physical construction, typing habit and so many factors decide the user's keystroke force. There is a need to adjust the hardness of the keyboard to suit the user's comfort.

**Solution by the invention**

Mochizuki discloses a method (US patent 5466901, Nov 95) of adjustable touch computer keyboard. Each key has a scissor like legs and compression coil springs, which push the key up. A slide mechanism is used to adjust the compression of the spring to vary the touch and feel of the key.

George English disclosed a method of adjustable force keyboard (patent 5879088, assigned to Key Tronic Corporation, Mar 99) that uses air pressure to adjust the softness of the keys. The air chamber is formed beneath the support plate extending air pressure to the keys. A manually adjustable air supply is connected to the air chamber to adjust the air pressure and thereby adjust the force required to actuate a key.

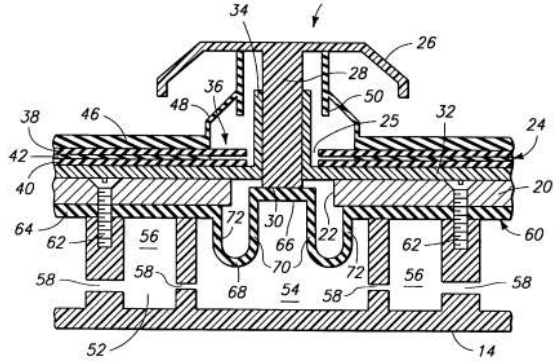

**TRIZ based analysis**

The tension of the keys should match with the striking force applied by the user **(desired result)**.

There are some attempts to make the key switch tensions adjustable. Some of them increases the cost of the keyboard, some other reduce the performance and life of the keyboard **(Limitations)**.

The invention of Mochizuki includes an adjustable mechanism for the key springs **(Principle-15: Dynamize)**.

The invention of English uses adjustable air pressure to adjust the keys **(Principle-29: Pneumatics and hydraulics)**.

**2.10 Low profile keyboard (Patent 5971637)**

**Background problem**

The keyboard of a portable computer needs to be small in size. The height of the keys even consumes a lot of storage space in a portable computer. It is necessary to reduce the height of the keys.

**Solution provided by the invention**

Satwinder Malhi, et al. invented a low profile key mechanism with full keystroke capability (patent 5971637, Assigned to Texas Instruments, Oct 99). The keyboard uses springs for the keystrokes. The springs give striking depth while the keyboard is active. While closed, the springs are disengaged from the guide mechanism to remain at lower height.

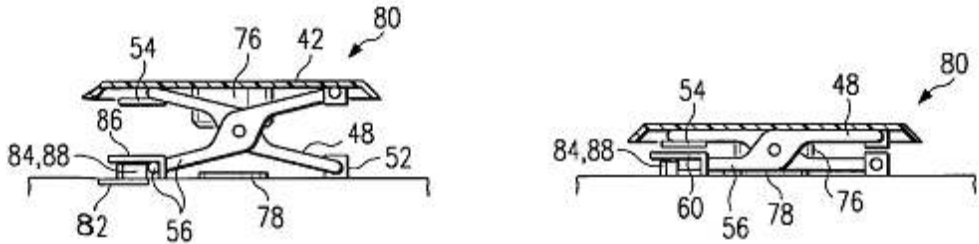

According to the invention, the key uses a lateral spring that can be positioned in two different states. The first state gives a depth to the keys to provide a resistive force against a keystroke. The second state is a lowered state with a reduced height. The state of the keys is changed by a guide mechanism.

Using this guide mechanism one can provide a full keystroke of three to five millimeters while in an active state and reduce its height to approximately two millimeters while in an inactive state. Since the spring moves laterally from the keys, no height is added to the keyboard by virtue of the spring movement.

**TRIZ based analysis**

The keys on the keyboard should be high to give a striking depth for typing comfort. But the keys should not be high to fit the keyboard inside the laptop **(Contradiction)**.

Change the position of the spring to lateral which will not require a height **(Principle-17: Another dimension)**.

The invention uses two states of the keyboard, a state with high depth for typing and a state with low depth for storing **(Principle-35: Parameter change)**.

## 3. Summary and conclusion

The different inventions on keyboard key switch mechanism intend to achieve various objectives. Some objectives achieved by the above inventions are as below:
- Simplify the mechanism for easy manufacturing.
- To use same rubber sheets for various key layouts thus reducing manufacturing problem
- Reduce cost of manufacturing
- Increase durability of the key switch mechanism
- Typing comfort by designing the key shape and size

- Reducing typing error or increasing reliability
- Reducing the height of the keys
- Adjusting softness of the keys according to user need
- Increase the strength and robustness of the key switches.

As key switches are very crucial to a keyboard, we can expect to see many more inventions improving various aspects of key switches in future.